\documentclass [12pt] {article}
\usepackage{a4}

  \def\pp{{\mathchoice
          {%displaystyle
              \kern 1pt%
              \raise 1pt
              \vbox{\hrule width5pt height0.4pt depth0pt
                    \kern -2pt
                    \hbox{\kern 2.3pt
                          \vrule width0.4pt height6pt depth0pt
                          }
                    \kern -2pt
                    \hrule width5pt height0.4pt depth0pt}%
                    \kern 1pt
           }
            {%textstyle
              \kern 1pt%
              \raise 1pt
              \vbox{\hrule width4.3pt height0.4pt depth0pt
                    \kern -1.8pt
                    \hbox{\kern 1.95pt
                          \vrule width0.4pt height5.4pt depth0pt
                          }
                    \kern -1.8pt
                    \hrule width4.3pt height0.4pt depth0pt}%
                    \kern 1pt
            }
            {%scriptstyle
              \kern 0.5pt%
              \raise 1pt
              \vbox{\hrule width4.0pt height0.3pt depth0pt
                    \kern -1.9pt  %[e]=0.15pt
                    \hbox{\kern 1.85pt
                          \vrule width0.3pt height5.7pt depth0pt
                          }
                    \kern -1.9pt
                    \hrule width4.0pt height0.3pt depth0pt}%
                    \kern 0.5pt
            }
            {%scriptscriptstyle
              \kern 0.5pt%
              \raise 1pt
              \vbox{\hrule width3.6pt height0.3pt depth0pt
                    \kern -1.5pt
                    \hbox{\kern 1.65pt
                          \vrule width0.3pt height4.5pt depth0pt
                          }
                    \kern -1.5pt
                    \hrule width3.6pt height0.3pt depth0pt}%
                    \kern 0.5pt%}
            }
        }}

  \def\mm{{\mathchoice
                       {%displaystyle
                             \kern 1pt
               \raise 1pt    \vbox{\hrule width5pt height0.4pt depth0pt
                                  \kern 2pt
                                  \hrule width5pt height0.4pt depth0pt}
                             \kern 1pt}
                       {%textstyle
                            \kern 1pt
               \raise 1pt \vbox{\hrule width4.3pt height0.4pt depth0pt
                                  \kern 1.8pt
                                  \hrule width4.3pt height0.4pt depth0pt}
                             \kern 1pt}
                       {%scriptstyle
                            \kern 0.5pt
               \raise 1pt
                            \vbox{\hrule width4.0pt height0.3pt depth0pt
                                  \kern 1.9pt
                                  \hrule width4.0pt height0.3pt depth0pt}
                            \kern 1pt}
                       {%scriptscriptstyle
                           \kern 0.5pt
             \raise 1pt  \vbox{\hrule width3.6pt height0.3pt depth0pt
                                  \kern 1.5pt
                                  \hrule width3.6pt height0.3pt depth0pt}
                           \kern 0.5pt}
                       }}

\catcode`@=11
\def\un#1{\relax\ifmmode\@@underline#1\else
        $\@@underline{\hbox{#1}}$\relax\fi}
\catcode`@=12

\let\du=\du                     % dot-under
\def\h{\eta}

\def\l{\lambda}
\def\m{\mu}

\def\p{\pi}
\def\q{\theta}
\def\r{\rho}

\def\x{\xi}

\def\D{\Delta}

\def\G{\Gamma}

\def\O{\Omega}

\def\Q{\Theta}

\def\ve{\varepsilon}

\def\bo{{\raise-.5ex\hbox{\large$\Box$}}}               % D'Alembertian
\def\pa{\partial}                                       % curly d
\def\de{\nabla}                                         % del
\def\TH{{\raise.2ex\hbox{$\displaystyle \bigodot$}\mskip-4.7mu \llap H \;}}
\def\face{{\raise.2ex\hbox{$\displaystyle \bigodot$}\mskip-2.2mu \llap {$\ddot
        \smile$}}}                                      % happy face
\def\abs#1{\left| #1\right|}                    % | |
\def\leftrightarrowfill{$\mathsurround=0pt \mathord\leftarrow \mkern-6mu
        \cleaders\hbox{$\mkern-2mu \mathord- \mkern-2mu$}\hfill
        \mkern-6mu \mathord\rightarrow$}
\def\dvec#1{\vbox{\ialign{##\crcr
        \leftrightarrowfill\crcr\noalign{\kern-1pt\nointerlineskip}
        $\hfil\displaystyle{#1}\hfil$\crcr}}}           % <--> accent
\def\frac#1#2{{\textstyle{#1\over\vphantom2\smash{\raise.20ex
        \hbox{$\scriptstyle{#2}$}}}}}                   % fraction
\def\sfrac#1#2{{\vphantom1\smash{\lower.5ex\hbox{\small$#1$}}\over
        \vphantom1\smash{\raise.4ex\hbox{\small$#2$}}}} % alternate fraction
\def\bfrac#1#2{{\vphantom1\smash{\lower.5ex\hbox{$#1$}}\over
        \vphantom1\smash{\raise.3ex\hbox{$#2$}}}}       % "
\def\afrac#1#2{{\vphantom1\smash{\lower.5ex\hbox{$#1$}}\over#2}}    % "
\def\[{\lfloor{\hskip 0.35pt}\!\!\!\lceil}
\def\]{\rfloor{\hskip 0.35pt}\!\!\!\rceil}

\def\du#1#2{_{#1}{}^{#2}}

\def\fracm#1#2{\hbox{\large{${\frac{{#1}}{{#2}}}$}}}

\def\un{\underline}
\def\fracmm#1#2{{{#1}\over{#2}}}

\def\low#1{{\raise -3pt\hbox{${\hskip 0.75pt}\!_{#1}$}}}

\newskip\humongous \humongous=0pt plus 1000pt minus 1000pt
\def\caja{\mathsurround=0pt}
\def\eqalign#1{\,\vcenter{\openup2\jot \caja
        \ialign{\strut \hfil$\displaystyle{##}$&$
        \displaystyle{{}##}$\hfil\crcr#1\crcr}}\,}
\newif\ifdtup

\def\pl#1#2#3{Phys.~Lett.~{\bf {#1}B} (19{#2}) #3}
\def\np#1#2#3{Nucl.~Phys.~{\bf B{#1}} (19{#2}) #3}

\parskip=\medskipamount                 % space between paragraphs (LaTeX)
\thicklines                         % thick straight lines for pictures (LaTeX)

\begin{document}
\thispagestyle{empty}

{\hbox to\hsize{
\vbox{\noindent July 2004 \hfill hep-th/0407160 }}}

\noindent
\vskip1.3cm
\begin{center}

{\Large\bf D-instanton Sums for Matter Hypermultiplets}

\vglue.2in

                  Sergei V. Ketov~\footnote{Supported in part by JSPS and 
Volkswagen Stiftung}

{\it          Department of Physics\\
           Tokyo Metropolitan University\\
          Hachioji, Tokyo 192--0397, Japan}
\vglue.1in
{\sl ketov@phys.metro-u.ac.jp}
\vglue.1in

                Osvaldo P. Santillan

{\it    Bogoljubov Laboratory of Theoretical Physics\\
         JINR, 141 980 Dubna, Moscow region, Russia}
\vglue.1in
               {\sl osvaldo@thsun1.jinr.ru}
\vglue.1in

{\rm and}
\vglue.1in
                  Andrei G. Zorin

{\it    Department of Physics, Moscow State University\\
            Vorobjovy Gory, 119899 Moscow, Russia}
\vglue.1in
                {\sl zrn@aport.ru}
\vglue.1in
\end{center}

\vglue.3in

\begin{center}
{\Large\bf Abstract}
\end{center}

\noindent
We calculate some non-perturbative (D-instanton) quantum corrections to the 
moduli space metric of several $(n>1)$ identical matter hypermultiplets for 
the type-IIA superstrings compactified on a Calabi-Yau threefold, near 
conifold singularities. We find a non-trivial deformation of the (real) 
$4n$-dimensional hypermultiplet moduli space metric due to the infinite 
number of D-instantons, under the assumption of $n$ tri-holomorphic commuting 
isometries of the metric, in the hyper-K\"ahler limit (i.e. in the absence of 
gravitational corrections). 

\newpage

\section{Introduction}

Derivation of non-perturbative `stringy' contributions to effective 
field theory is one of the major problems in string theory towards its 
phenomenological applications. In the absence of fundamental formulation of
non-perturbative string/M-theory, any explicit example of summing up some
non-perturbative corrections may shed light on other cases as well. The 
non-perturbative low-energy effective action for hypermultiplets in the 
{\it Calabi-Yau} (CY) compactified type-IIA superstings is a good place for 
such calculations, because the relevant moduli space metric in the effective 
N=2 supergravity is affected by quantum corrections due to D-instantons and 
five-brane instantons. The former originate from the 
(Euclidean) D2-branes  wrapped about certain (special Lagrangian) 3-cycles in
CY threefold, whereas the latter come from the (NS-NS, Euclidean) five-branes 
wrapped about the entire CY \cite{bbs}.

With this motivation in mind, most attention in the past was devoted to the 
so-called {\it universal hypermultiplet} (UH) present in any CY 
compactification \cite{har,myu,ant,van}. The UH contains dilaton amongst
its bosonic physical components, while it is essentially gravitational in 
nature (i.e. gravitational corrections cannot be ignored for UH). 

The D-instanton quantum corrections to the quantum moduli space metric of a 
{\it single matter hypermultiplet} for the CY-compactified type IIA 
superstrings near a conifold singularity were calculated by Ooguri and Vafa 
\cite{ov}. They found the unique solution consistent with N=2 {\it rigid} 
supersymmetry and 
toric isometry. The solution \cite{ov} was interpreted as the infinite 
D-instanton sum coming from multiple wrappings of the Euclidean D2-branes 
around the vanishing cycle \cite{ov}. The Ooguri-Vafa solution is given by the 
hyper-K\"ahler metric in the limit of flat four-dimensional spacetime, i.e. 
when N=2 supergravity decouples. The hyper-K\"ahler solution \cite{ov} was 
lifted to a quaternionic solution in curved spacetime of N=2 supergravity in 
ref.~\cite{mym}.

With almost all known cases being limited to a single hypermultiplet, it is
quite natural to turn attention to many hypermultiplets. A generic CY 
compactification is well known to yield $n={\rm dim}\,H^{2,1}+1$ 
hypermultiplets, where ${\rm dim}\,H^{2,1}$ is a Hodge number of CY. 
Unfortunately, the geometrical constraints imposed on hypermultiplets by N=2
supersymmetry are weaker for several hypermultiplets when compared to only one.
This may, however, be compensated by requiring enough unbroken symmetries. Our
calculations in this Letter are performed in the hyper-K\"ahler limit when 
both N=2 supergravity and UH are switched off. The five-brane instanton
corrections are weighted by powers of $e^{-1/g^2}$, whereas that of 
D-instantons are weighted by powers of  $e^{-1/g}$, where $g$ is string 
couping \cite{wit}. Hence, when the string coupling is sufficiently large (as
is the case in our study), the five-brane instantons can be suppressed against 
the D-instantons. As a final simplification, we assume the existence of $n$ 
tri-holomorphic (i.e. commuting with N=2 supersymmetry) isometries of the 
D-instanton corrected quantum moduli space metric of $n$ (identical) matter 
hypermultiplets.

Our paper is organized as follows. In sect.~2 we birefly review the Ooguri-Vafa
solution \cite{ov} for a single matter hypermutliplet. In sect.~3 we introduce
 some mathematical tools needed to set up the framework to our calculations. 
In sect.~4 we give the results of our calculations for the metric. Sect.~5 is
our conclusion.
 
\section{Ooguri-Vafa solution}

The {\it Ooguri-Vafa} (OV) solution \cite{ov} describes the D-instanton 
corrected moduli space metric of a single matter hypermultiplet in type-IIA 
superstrings compactified on a Calabi-Yau threefold of Hodge number 
${\rm dim}H^{2,1}=1$, when {\it both} N=2 supergravity {\it and} UH are 
switched off, while five-brane instantons are suppressed. The matter 
hypermultiplet low-energy effective action in that limit is given by the 
four-dimensional N=2 supersymmetric non-linear sigma-model that has the 
four-dimensional OV metric in its target space.

Rigid (or global) N=2 supersymmetry of the non-linear sigma-model requires a 
hyper-K\"ahler metric in its target space \cite{fal,book}. The OV metric has a 
(toric) $U(1)\times U(1)$ isometry by construction \cite{ov}. There always 
exist a linear combination of two commuting abelian isometries that is 
tri-holomorphic, i.e. it commutes with N=2 rigid supersymmetry \cite{gib}.

Given any four-dimensional hyper-K\"ahler metric with a tri-holomorphic 
isometry $\pa_t$, it can always be written down in the standard 
(Gibbons-Hawking) form \cite{gh},
$$ ds^2_{\rm GH}= \fracmm{1}{V}(dt+\hat{\Q})^2 +V(dx^2+dy^2+dz^2)~, 
\eqno(1)$$
that is governed by {\it linear} equations,
$$ \D V=\vec{\nabla}{}^2V\equiv \left( \fracmm{\pa^2}{\pa x^2}+
\fracmm{\pa^2}{\pa y^2}
+\fracmm{\pa^2}{\pa z^2}\right)V=0~,\quad {\rm almost~~everywhere},\eqno(2)$$
and
$$\vec{\de}V+\vec{\de}\times\vec{\Q}=0~.\eqno(3)$$
The one-form $\hat{\Q}=\Q_1dx +\Q_2dy+\Q_3dz$ is fixed by the 
`monopole equation' (3) in terms of the real scalar potential $V(x,y,z)$. 
Equation (2) means that the function $V$ is harmonic (away from possible
 isolated singularities) in three Euclidean dimensions ${\bf R}^3$. The 
singularities are associated with the positions of D-instantons.

Given extra $U(1)$ isometry, after being rewritten in the cylindrical 
coordinates ($\r=\sqrt{x^2+y^2},~\theta=\arctan(y/x),~\h=z)$, the 
hyper-K\"ahler potential $V(\r,\q,\h)$ becomes independent upon $\theta$. 
Equation (1) was used by Ooguri and Vafa \cite{ov} in their analysis of the 
matter hypermultiplet moduli space near a conifold singularity. The conifold 
singularity arises in the limit of the vanishing CY period, 
$$  \int_{\cal C}\O\to 0~~,\eqno(4)$$
where the CY holomorphic (nowhere vanishing) 3-form $\O$ is integrated over a 
non-trivial 3-cycle ${\cal C}$ of CY. The powerful singularity theory 
\cite{sin}  can then be applied to study the universal behaviour of the 
hypermultiplet moduli space near the conifold limit, by resolving the 
singularity.

In the context of the CY compactification of type IIA superstrings, the
coordinate $\r$ represents the `size' of the CY cycle ${\cal C}$ or, 
equivalently, the action of the D-instanton originating from the Euclidean 
D2-brane wrapped about the cycle ${\cal C}$. The physical interpretation of 
the $\h$ coordinate is just the expectation value of another (RR-type) 
hypermultiplet scalar. The cycle ${\cal C}$ can be replaced by a sphere $S^3$
for our purposes, since the D2-branes only probe the overall size of 
${\cal C}$.

The OV potential $V$ is {\it periodic} in the RR-coordinate $\h$ since 
the D-brane charges are quantized \cite{bbs}. We normalize the corresponding 
period to be $1$, as in ref.~\cite{ov}. The Euclidean D2-branes wrapped $m$ 
times around the sphere $S^3$ couple to the RR  expectation value on $S^3$ and
thus should produce additive contributions to $V$, with the factor of 
$\exp(2\p im\h)$ each.

In the {\it classical} hyper-K\"ahler limit, when both N=2 supergravity and 
all the D-instanton contributions are suppressed, the potential $V(\r,\h)$ of
a single matter hypermultiplet cannot depend upon $\h$ since there is no 
perturbative superstring state with a non-vanishing RR charge. Accordingly, 
the classical pre-potential $V(\r)$ can only be the Green function of the 
two-dimensional Laplace operator, i.e. 
$$ V_{\rm classical} = -\fracmm{1}{2\p}\log\r + {\rm const.}~,\eqno(5)$$
whose normalization is also in agreement with ref.~\cite{ov}. 

The calculation of ref.~\cite{ov} to determine the exact D-instanton 
contributions to the hyper-K\"ahler potential $V$ is based on the idea 
\cite{bbs} that the D-instantons should resolve the singularity of the 
classical hypermultiplet moduli space metric at $\r=0$. A similar situation 
arises in the standard (Seiberg-Witten) theory of a quantized N=2 vector 
multiplet (see e.g., ref.~\cite{book} for a review).

Equation (2) formally defines the electrostatic potential $V$ of electric 
charges of unit charge in the Euclidean upper half-plane $(\r,\h)$, $\r>0$, 
which are distributed along the axis $\r=0$ in each point $\h=n\in {\bf Z}$, 
while there are no two charges at the same point \cite{ov}. A solution to 
eq.~(2) obeying all these conditions is unique, 
$$ V\low{\rm OV}(\r,\h)= \fracmm{1}{4\p} \sum^{+\infty}_{n=-\infty}\left(
\fracmm{1}{\sqrt{\r^2+ (\h-n)^2}}-\fracmm{1}{\abs{n}}\right)+{\rm const.}
\eqno(6)$$
After Poisson resummation eq.~(6) takes the desired form of singularity 
resolution \cite{ov}:
$$ V\low{\rm OV}(\r,\h)=\fracmm{1}{4\p} \log\left( \fracmm{\m^2}{\r^2}\right)+
\sum_{m\neq 0}\fracmm{1}{2\p}e^{2\p im\h}\,K_0\left(2\p \abs{m}\r\right)~,
\eqno(7)$$
where the modified Bessel function $K_0$ of the 3rd kind has been introduced,
$$ K_s(z)=\fracm{1}{2}\int^{+\infty}_0\fracmm{dt}{t^{s-1}}\exp\left[
-\,\fracmm{z}{2}\left( t+\fracmm{1}{t}\right)\right]~,\eqno(8)$$
valid for all Re\,$z>0$ and  Re\,$s>0$, while $\m$ is a constant (modulus).

Inserting the standard asymptotical expansion of the Bessel function $K_0$ 
near $\r=\infty$ into eq.~(7) yields \cite{ov}
$$\eqalign{
 V\low{\rm OV}(\r,\h)~=~&\fracmm{1}{4\p} \log 
\left( \fracmm{\m^2}{\r^2}\right) +
\sum_{m=1}^{\infty} \exp \left(-2\p m\r\right)
\cos(2\p m\h)\times\cr
~& \times \sum_{n=0}^{\infty}\fracmm{\G(n+\fracm{1}{2})}{\sqrt{\p}n!
\G(-n+\fracm{1}{2})}\left(\fracmm{1}{4\p m\r}
\right)^{n+\frac{1}{2}}~~~.\cr}\eqno(9)$$

The string coupling constant $g$ can be easily reintroduced into eq.~(9) by a 
substitution $\r\to\r/g$. The factors of $\exp{(-2\p m\r/g)}$ in eq.~(9) are 
the contributions due to the multiple D-instantons \cite{ov}. 

The OV potential (6) is given by a (regularized) T-sum over the
 T-duality transformations, $\h\to \h+1$, being applied to the fundamental
solution 
$V_0\equiv\fracmm{1}{4\p r}\equiv \fracmm{1}{4\p\sqrt{\r^2+\h^2}}$ of eq.~(2),
$$ V\low{\rm OV}(\r,\h)=A + \sum_{\rm T}V_0(\r,\h)= A + \sum_{\rm T} 
\fracmm{1}{4\p\sqrt{\r^2+\h^2}}~~~,\eqno(10)$$
where $A$ is a constant. The fundamental solution $V_0(\r,\h)$ is just the 
Green function of the three-dimensional Laplace operator $\D$ in eq.~(2).

\section{Pedersen-Poon Ansatz}

There exists a natural generalization of the Gibbons-Hawking Ansatz (1) to
higher-dimensional toric hyper-K\"ahler spaces, known as the {\it 
Pedersen-Poon} (PP) metric \cite{pp}. Namely, given $n$ commuting
tri-holomorphic isometries of a (real) $4n$-dimensional hyper-K\"ahler space,
there exists a coordinate system $(x^i_a,t_i)$, with $i,j=1,2,\ldots,n$ and
$a,b,c=1,2,3$, where the hyper-K\"ahler metric takes the form \cite{pp}:
$$ ds^2= U_{ij}dx^i\cdot dx^j+ U^{ij}(dt_i+A_i)(dt_j +A_j)~~.\eqno(11)$$
Here the dot means summation over the $a$-type indices, all metric 
components are supposed to be independent upon all $t_i$ (thus reflecting the 
existence of $n$ isometries), $U^{ij}=(U_{ij})^{-1}$,  while 
$U_i=(U_{i1},\ldots,U_{in})$ and $A_i$ are to be solutions to the generalized
 monopole (or BPS) equations:
$$ {\cal R}_{x^i_ax^j_b}=\ve_{abc}\de_{x^i_c}U_j \quad
{\rm and} \quad \de_{x^i_a}U_j=\de_{x^j_a}U_i~~,\eqno(12)$$
where the field strength ${\cal R}$ of the gauge fields $A$ has been 
introduced. The PP metric (11) can be completely specified by its real 
PP-potential $F(x,w,\bar{w})$ that generically depends upon $3n$ variables,
$$  x^j=x^j_3~,\qquad w^j=\fracmm{x^j_1+ix^j_2}{2}~,\qquad 
\bar{w}^j=\fracmm{x^j_1-ix^j_2}{2}~,\eqno(13)$$
because the BPS equations (12) allow a solution \cite{pp}
$$  U_{ij}=F_{x^ix^j}\quad {\rm and}\quad
A_j=i\left(F_{w^kx^j}dw^k-F_{\bar{w}^kx^j}d\bar{w}^k\right)~,\eqno(14)$$
provided $F$ itself obeys (almost everywhere) a linear Laplace-like equation 
\cite{pp}
$$ F_{x^ix^j} +F_{w^i\bar{w}^j}=0~.\eqno(15)$$
The subscripts of $F$ denote partial differentiation with respect to the
given variables. 

Equation (15) is apparently the multi-dimensional hyper-K\"ahler 
generalization of eq.~(2). The existence of the PP-potential $F$ essentially 
amounts to integrability (or linearization) of the BPS equations (12), because
the master equation (15) is linear, whose solutions are not difficult to find.
For instance, a general solution to eq.~(15) can be written down as the
contour $(C)$ integral of an arbitrary potential $G(\h^j(\x),\x)$  \cite{con},
$$ F= {\rm Re} \oint_C \fracmm{d\x}{2\p i\x}G(\h^j(\x),\x)~,\quad {\rm where}
\quad \h^j(\x) = \bar{w}^j+x^j\x-w^j\x^2~~.\eqno(16)$$  
   
\section{D-instanton sums}

Our technical assumptions are essentially the same as in ref.~\cite{ov}, 
namely,
\begin{itemize}
\item periodicity in all $x^i$ with period $1$, due to the D-brane charge 
quantization,
\item the classical potential $F$ near a CY conifold singularity should have a
logarithmic behaviour (elliptic fibration), being independent upon all $x^i$ 
(when all $w^i\to\infty),$
\item the classical singularity of the metric should be removable, i.e. an
exact metric should be complete (or non-singular),
\item the metric should be symmetric under the permutation group of $n$ sets
of hypermultipet coordinates $(x^j,w^j,\bar{w}^j)$, where $j=1,2,\ldots,n$.  
\end{itemize}
The last assumption means that we only consider {\it identical} matter 
hypermultiplets. Together with our main assumption about $n$ tri-holomorphic  
isometries (see e.g., our Abstract and sect.~3), it is going to lead us to
 an explicit solution.

We first confine ourselves to the case of {\it two} identical matter
hypermultiplets $(n=2)$, and then quote our result for an arbitrary $n>2$. 
The problem amounts to finding a solution to the master equations (15) subject
to the conditions above. The known OV solution 
$V\low{\rm OV}(\abs{w},x)\equiv V\low{\rm OV}(x,w)$, defined by eqs.~(6) or 
(7), can be used to introduce another function $F(x,w)$ as a solution to two 
equations 
$$ F_{xx}=-F_{w\bar{w}}=V_{\rm OV}(x,w)~.\eqno(17)$$
We do not need an explicit solution to eq.~(17) in what wollows. Since 
eqs.~(15) are linear, the superposition principle applies to their solutions. 
It is now straightforward to verify that there exist a `trivial' solution 
to eqs.~(15), given by a PP potential 
$$ F_0= c_0 \left[F(x_1,w_1) + F(x_2,w_2)\right]~~,\eqno(18)$$ 
where $c_0$ is a real constant. A particular non-trivial solution  in the 
$n=2$ case is given by
$$ F_{\rm mixed} = c_{+}F(x_1+x_2,w_1+w_2) + 
c_{-}F(x_1-x_2,w_1-w_2)~~,\eqno(19)$$
where $c_{\pm}$ are two other real constants. Though the linear partial
differential equation (15) has many other solutions, we argue at the end of 
this section that the most general solution, satisfying all our requirements, 
is actually given by
$$ F=F_0+F_{\rm mixed}~~.\eqno(20)$$
To this end, we continue with the particular solution (20).

\newpage

To write down our result for the hyper-K\"ahler moduli space metric of two 
matter hypermultiplets, we merely need the second derivatives of the PP
potential, because of eq.~(14). Using eqs.~(6), (14), (17), (18), (19) and 
 (20), we find
$$ \eqalign{
4\p U_{11} ~=~ & ~~c_+\sum^{+\infty}_{n=-\infty}\left(\fracmm{1}{\sqrt{
(x_1+x_2-n)^2 +(w_1+w_2)(\bar{w}_1+\bar{w}_2)/\l^2}}-\fracmm{1}{n}\right)\cr
 ~ & +c_-\sum^{+\infty}_{n=-\infty}\left(\fracmm{1}{\sqrt{
(x_1-x_2-n)^2 +(w_1-w_2)(\bar{w}_1-\bar{w}_2)/\l^2}}-\fracmm{1}{n}\right)\cr
~ & +c_0\sum^{+\infty}_{n=-\infty}\left(\fracmm{1}{\sqrt{
(x_1-n)^2 +w_1\bar{w}_1/\l^2}}-\fracmm{1}{n}\right)~~,\cr}\eqno(21)$$
with a modulus parameter $\l$. The component $U_{22}$ has the the form as 
 that of eq.~(21), but the indices $1$ and $2$ have to be exchanged. The 
remaining two components of the matrix $U$ in eq.~(11) are given by
$$ \eqalign{
4\p U_{12}=4\p U_{21} = & ~~c_+\sum^{+\infty}_{n=-\infty}\left(
\fracmm{1}{\sqrt{(x_1+x_2-n)^2 +(w_1+w_2)(\bar{w}_1+\bar{w}_2)/\l^2}}-
\fracmm{1}{n}\right)\cr
  & + c_-\sum^{+\infty}_{n=-\infty}\left(\fracmm{1}{\sqrt{
(x_1-x_2-n)^2 +(w_1-w_2)(\bar{w}_1-\bar{w}_2)/\l^2}}-\fracmm{1}{n}\right)~.\cr}
\eqno(22)$$
The $A$-field components in the PP metric (11) are given by the second 
equation (14). 

The physical interpretation of our solution as the infinite D-instanton sum is
evident after rewriting it as an asymptotical expansion like that of eq.~(9),
by using eqs.~(7) and (8), now being applied to the two-hypermultiplet metric 
components (21) and (22). In particular, their classical behavior is given by
$$U_{11} ~\sim~ \fracmm{c_+}{4\p}\ln\fracmm{\l^2}{\abs{w_1+w_2}^2}+
\fracmm{c_-}{4\p}\ln\fracmm{\l^2}{\abs{w_1-w_2}^2}+
\fracmm{c_0}{4\p}\ln\fracmm{\l^2}{\abs{w_1}^2}~~~,\eqno(23)$$
and similarly for $U_{22}$ after exchanging the indices $1$ and $2$, and 
$U_{12}$ after dropping the last term in eq.~(23).  
 
As regards the case of an arbitrary $n>2$, the easiest picture arises in terms
 of the PP potential $F$ --- see sect.~3. We define this potential as a linear
 combination of $F(x_1+x_2+\ldots +x_n, w_1+w_2+\ldots +w_n)$,  
$F(-x_1+x_2+\ldots +x_n, -w_1+w_2+\ldots +w_n)$, $\ldots$,
$F(-x_1-x_2+\ldots -x_n, -w_1-w_2-\ldots -w_n)$, $F(x_1,w_1)$, $F(x_2,w_2)$,
$\ldots$ and $F(x_n,w_n)$, each one being construcuted out of the OV potential,
as in eq.~(17). Next, we define the metric components of the matrix $U_{ij}$ 
in eq.~(11) by the second derivatives of the total PP potential $F$, as in
eq.~(14). The rest is fully straightforward, {\it cf.} eqs.~(21) and (22). 

\newpage

Finally, some comments about the uniqueness of our solution (20) are in order. 
First, we notice that all functions $U^{ij}$ also obey eq.~(15) by our
construction (14). Given any other non-trivial solution, it should lead 
(after  Poisson resummation) to {\it the same} classical behaviour (23), while
it should also  reduce to the OV solution (sect.~2) for a single 
hypermultiplet.  According to the Poisson resummation formula,
$$ \sum^{\infty}_{n=0}f(n) = \sum^{\infty}_{n=0}\tilde{f}(n)~,\quad{\rm where}
\quad \tilde{f}(u)=\int^{+\infty}_{-\infty}dvf(v)e^{2\p iuv}~~,\eqno(24)$$
the logarithmic terms can only come from the $n=0$ term in the sum, e.g.
when using eq.~(6), choosing $n=v$ in the Fourier integral (24), and then
taking $u=0$ after the Fourier transform. 

Hence, a difference between ours and any other solution to the potentials $F$
or $U^{ij}$ cannot contribute to eq.~(23).  Since eq.~(23) is $x$-independent,
we first investigated the class of $x$-independent real potentials $F$, 
without any reference to the metric (11). By solving eq.~(15) in this case we 
found an unique solution subject to our symmetries, namely, the one given by
eq.~(23). Hence, the anticipated uniqueness of our solution boils down to the 
uniqueness of the periodic (of period $1$) extension of a given $w$-dependent 
solution to a solution depending upon both $w$ and $x$, and having a generic 
form $\sum_n G(x+n,w)$ with some basic function $G$. It seems to be quite 
plausible that the square root in eq.~(21) is the {\it only} basic function to
 yield the logarithmic terms out of the $n=0$ term in the D-istanton sum.
  
\section{Conclusion}

In our final results (21) and (22) the quantum corrections to the classical 
(logarithmic) terms (23) are exponentially suppressed by the D-instanton 
factors (in the semiclassical description). Unlike the case of a single
matter hypermultiplet (sect.~2), the multi-hypermultiplet moduli space metric
is also sensitive to the phases of $w_i$ (i.e. not only to their absolute 
values). There are no perturbative (type IIA superstring) corrections, while 
all D-instanton numbers contribute to the solution.

Our results can also be applied to an explicit construction of {\it 
quaternionic} metrics in real $4(n-1)$ dimensions out of known hyper-K\"ahler 
metrics in real $4n$ dimensions, which is of particular importance to a 
description of D-instantons in N=2 supergravity \cite{swan}. For instance, 
the four-dimensional quaternionic manifolds with toric isometry can be fully 
classified \cite{cp} in terms of the PP-potential $F$ obeying eq.~(15). 

It would be also interesting to connect our results to perturbative superstring
calculations in the D-instanton background, which is still a largely unsolved
problem.

\end{document}